\def \kms{\rm{km}~$\rm{s}^{-1}$}

\def \cm{~\rm{cm}}
\def \s{~\rm{s}}
\def \km{~\rm{km}}
\def \kms{$~\rm{km}~{\rm s}^{-1}$}

\def \K{~\rm{K}}

\def \erg{~\rm{erg}}

\def \yr{~\rm{yr}}

\def \kev{~\rm{keV}}
\documentclass[12pt,preprint]{aastex}
\usepackage{natbib}
\usepackage{epsfig}
\begin{document}   

\title{X-Ray Emission by A Shocked Fast Wind from the Central Stars of Planetary Nebulae}

\author{Muhammad Akashi, Noam Soker, and Ehud Behar}
\affil{Department of Physics, Technion$-$Israel Institute of
Technology, Haifa 32000, Israel, Email:
akashi@physics.technion.ac.il; soker@physics.technion.ac.il;
behar@physics.technion.ac.il}

%

\begin{abstract}
We calculate the X-ray emission from the shocked fast wind blown
by the central stars of planetary nebulae (PNs) and compare with
observations. Using spherically symmetric self similar solutions,
we calculate the flow structure and X-ray temperature for a fast
wind slamming into a previously ejected slow wind. We find that
the observed X-ray emission of six PNs can be accounted for by
shocked wind segments that were expelled during the early PN
phase, if the fast wind speed is moderate, $v_2 \sim$
400-600~\kms, and the mass loss rate is a few times $10^{-7}
M_\odot \yr^{-1}$. We find, as proposed previously, that the
morphology of the X-ray emission is in the form of a narrow ring
inner to the optical bright part of the nebula. The bipolar X-ray
morphology of several observed PNs, which indicates an important
role of jets rather than a spherical fast wind, cannot be
explained by the flow studied here.

\end{abstract}

{\it Subject headings:} Subject headings: stars: mass loss $—$
stars: winds, outflows $—$ planetary nebulae: X-ray $—$ X-rays:
ISM

\section{INTRODUCTION}

The sample of well observed planetary nebulae (PNs) with extended
spatially resolved X-ray structures is large enough to start
answering open questions on the shaping of and the emission from
PNe. This sample includes PNs observed by the {\it Chandra} X-ray
Observatory (CXO): BD~+30$^\circ$3639 (Kastner et al.\ 2000;
Arnaud et al.\ 1996 detected X-rays in this PN with ASCA), NGC
7027, (Kastner, et al.\ 2001), NGC 6543 (Chu et al.\ 2001), Henize
3-1475 (Sahai et al. 2003), Menzel 3 (Kastner et al.\ 2003), as
well as PNs observed with {\it XMM-Newton}: NGC 7009 (Guerrero et
al.\ 2002), NGC 2392 (Guerrero et al.\ 2005), and MGC 7026
(Gruendl et al. 2004). For some other PNs no X-ray emission has
been detected: M1-16 (Soker \& Kastner 2003), NGC 7293 (Guerrero
et al.\ 2001), Hen 2-99 (Kastner et al.\ 2005), and NGC 2346
(Gruendl et al. 2004). It should be pointed out that X-ray
emission is expected from all PNs, but many PNs emit X-rays below
the current detection limit. In NGC 40, only a careful analysis,
looking selectively at photons in a specific energy band, has
yielded a positive detection (Kastner et al. 2005). The main
question now as formulated by Soker \& Kastner (2003) is: What is
the (astro)physical origin of the hot gas in the extended X-ray
emission region. There are several plausible answers. The X-ray
emitting gas may result from shocked fast wind segments that were
expelled by the central star during the early PN phase or late
post-asymptotic giant branch (AGB) phase, when the wind speed was
moderate, $\sim 500 \km \s^{-1}$. Alternatively, the X-ray
emitting gas may result from a collimated fast wind (CFW; or jets
if the wind is well collimated) blown in conjunction with the
companion to the central star during the late AGB phase or early
post-AGB phase. Soker \& Kastner (2003) conclude that in order to
account for the observed X-ray temperature and luminosity, both
the evolution of the wind from the central star and the adiabatic
cooling of the post-shocked wind's material must be considered.

Extended X-ray emission in PNs was expected based on the presence
of a fast wind driven by the central star during the late post-AGB
phase and early PN phase, as the interaction between this fast
wind and slower-moving material, ejected when the central star was
still an AGB star, should lead to energetic shocks. The same
wind-wind interaction plays some role in shaping many PNs (e.g.,
Balick \& Frank 2002, and references therein). However, CFWs (or
jets) must play a more significant role in the shaping of many PNs
(but not all PNs) than the spherical fast wind from the central
star plays (Soker 2004). Hence, the processes behind the extended
X-ray emission are tied to the shaping processes of PNs (Kastner
et al.\ 2003), although not all shaping processes will lead to
X-ray emission. X-ray-emitting regions of some PNs are quite
asymmetric. This asymmetry results in large part from extinction
(Kastner et al.\ 2003). However, extinction by itself cannot
account for all of the observed X-ray asymmetries and structural
irregularities, and other processes should be considered (Kastner
et al.\ 2003).

These findings, though, are not enough by themselves to reveal the
origin of the X-ray emitting gas, whether it is from a CFW blown
by a companion during the AGB and/or early post-AGB phase, or
whether the spherical fast wind blown by the central star during
the late post-AGB phase is the source of the X-ray emitting gas.
Theoretical calculations of the X-rays expected in these two
cases, and quantitative comparison with observations are required
for that purpose. It is very likely that both processes contribute
to a certain degree in different PNs. In the present paper we
study the expected contribution from the fast spherical wind blown
by the central star. We use the self-similar solution of Chevalier
\& Imamura (1983; hereafter CI83) and estimate the conditions
under which the wind blown by the central star can explain the
observed X-ray properties of PNs. In Section 2, we present the
theoretical method based on the self similar solution as well as
the physical parameters used to determine the X-ray emission. The
results of the calculations and the comparison with observations
are discussed in Sections 3 and 4, respectively. A summary of our
main findings is given in Section 5.

\section{THEORETICAL METHOD}

\subsection{Self Similar Solution}

CI83 present a self-similar solution to the following flow
structure. A spherically symmetric slow wind with a constant mass
loss rate $\dot M_1$ and a constant velocity $v_1$ fills the space
around the center $r=0$ where the star resides(see Fig.~1). At
time $t=0$ the slow wind ceases and a spherically symmetric fast
wind of mass loss rate $\dot M_2$ and a constant velocity $v_2$
expands and collides with the slow wind. The slow and fast wind do
not mix, but rather are separated by an evolving spherical surface
termed the contact discontinuity corresponding to a radius
$R_c(t)$ (see Fig.~2 by Volk~\& Kwok 1985 and fig. 1 here). A
spherical shock wave runs outwards into the slow wind at radius
$R_1(t)>R_c(t)$. At the same time, a spherical inward (reverse)
shock wave propagates through the fast wind down to radius $R_2$,
thus heating the region $R_2 < r < R_c$. This region is termed the
hot bubble. In the self similar solution, there are no imposed
characteristic length scales or time scales, hence the evolution
with time and distance depends only on the ratio $r/t$.

We are interested in the contribution of the shocked spherically
symmetric fast wind blown by the central star during the late
post-AGB and early PN phases to the X-ray emission. Therefore, we
examine only the region inward to the contact discontinuity
$R_2<r<R_c$. The solution in that region is given by specifying
two dimensionless parameters (CI83): (1) the ratio $b_2=v_2/s_2$,
where $s_2=\dot R_2(t)$ is the reverse shock speed; and (2) the
Mach number, which is the ratio $M_{2}=(v_2-s_2)/c_2$, where $c_2$
is the sound speed in the pre-shock fast wind gas. For the
appropriate parameters in the present case of $v_2 \ge 300 \km
\s^{-1}$ and central star effective temperature of $\sim 30,000
\K$, $v_2 >> c_2$ and the Mach number of the fast wind is very
large, practically infinite in the self-similar regime. For large
Mach numbers the value of $b_2$ is determined by the two ratios
$v_1/v_2$ and $\dot M_1/\dot M_2$ (Fig.~6. of CI83).

For large Mach numbers pressure equilibration is efficient and
thus the pressure is fairly constant throughout the hot bubble
(Fig.~4 of CI83 and Figs.~2 and 3 below). By assuming constant
pressure inside the hot bubble, Volk \& Kwok (1985) were able to
solved for a time delay $\tau_0 > 0$ between the end of the slow
wind and the beginning of the fast wind. Because the wind is
continuous and there is no real delay between the slow and fast
winds, we prefer to use the self similar solution to determine the
exact pressure profile inside the hot bubble. For the present
study, it is adequate to use the contact discontinuity speed as
given by Volk \& Kwok (1985; their Eq.~4):
\begin{equation}
\left(\frac{v_c}{v_1}\right)^3-2\left(\frac{v_c}{v_1}\right)^2+
\left(\frac{v_c}{v_1}\right)= \frac{1}{3} \frac {\dot M_2
v_2^2}{\dot M_1 v_1^2}. \label{vc}
\end{equation}

For the radius of the contact discontinuity we use the
approximation $R_c=v_c t$, where $t$ is the age of the fast wind
assumed to be also the age of the nebula. This time enters as an
additional parameter into our model. It is easy to find the
location of the reverse shock by equating the pressure in the hot
bubble given by Volk \& Kwok (1985) to the pre-shock ram pressure
of the fast wind. This gives for $R_2$:
\begin{equation}
\eta \equiv \frac{R_2}{R_c} = \left(\frac{\dot M_2 v_2}{\dot M_1
v_1}\right)^{1/2} \frac{v_1}{v_c-v_1} \label{r2}
\end{equation}

In summary, our calculations require the parameters $\dot M_1$,
$\dot M_2$, $v_1$, $v_2$, and $t$ as input. The self similar
solution then provides the density and temperature (and pressure)
profiles from $R_2$ to $R_c$. In the present study, we mainly vary
$v_2$ and $\dot M_2$, on which the X-ray properties are strongly
dependent with $v_1=10 \km \s^{-1}$ kept constant for all cases.

\subsection{Radiative Cooling}
The shocked fast wind loses energy and cools down as it emits
X-ray and UV radiation. This cooling must be considered in a full
numerical study as it cannot be dealt with in the self similar
formalism. In the present study, we do include cooling, but in a
rather crude way. We remove from the calculation of the X-ray
luminosity gas shells whose radiative cooling times are shorter
than the age of the nebula $t$.

The cooling (energy loss) rate per unit volume per unit time is
$\Lambda n_e n_p$, where $n_e$ and $n_p$ are the electron and
proton number densities, respectively. In the temperature range $2
\times 10^5$ K $\lesssim$ T $\lesssim 2 \times 10^7$ K, the
cooling function $\Lambda$ is approximately 10$^{-22} (T/10^6
\mathrm {K})^{-1/2} \erg \cm^3 \s^{-1}$, which is an approximation
to Fig.~6 of Gaetz et al.\ (1988). As the gas cools its cooling
time decreases quickly. Soker \& Kastner (2003) take this into
account, and write for the cooling time
\begin{equation}
\tau_{\rm cool} \simeq k_{\rm cool} \frac {n k T} {\Lambda n_e
n_p} = k_{\rm cool} \frac {P}{\Lambda n_e n_p}, \label{tcool1}
\end{equation}
where $n$ is the total number density including protons, electrons
and all other atoms and ions. At constant pressure and slow
variation with temperature of $T/\Lambda$, $k_{\rm cool} \simeq
5/2$. However, here $T/\Lambda \propto T^{3/2}$, and the cooling
time is somewhat shorter. Soker \& Kastner (2003) included also
adiabatic cooling in $k_{\rm cool}$, and for their crude estimate
took $k_{\rm cool}=1$. As adiabatic cooling is already included in
the self-similar solution, we take $k_{\rm cool}=5/2$.
Substituting for the typical values, Soker \& Kastner (2003) find
\begin {equation}
\tau_{\rm cool} \simeq
400  
\left( \frac {T} {10^6 \K}  \right)^{5/2} \left( \frac {v_{\rm c}}
{20 \km \s^{-1}}  \right)^2 \left( \frac {t} {500 \yr}  \right)^2
\left( \frac {\dot M_2 v_2} {3 \times 10^{-5} M_\odot \yr^{-1} \km
\s^{-1}}
   \right)^{-1}
\left( \frac {\eta} {0.3}  \right)^2 \yr. \label{tcool2}
\end{equation}
For a parcel of gas to stay in the X-ray emitting temperature for
a time $t$ the cooling time must be $\tau_{\rm cool} \ga t$, which
for the scaling used in equation (\ref{tcool2}) is equivalent to
\begin{equation}
T_{\rm min} \ga 10^6  
\left( \frac {v_{\rm c}} {20 \km \s^{-1}}  \right)^{-0.8} \left(
\frac {t} {500 \yr}  \right)^{-0.4} \left( \frac {\dot M_2 v_2} {3
\times 10^{-5} M_\odot \yr^{-1} \km \s^{-1}}
   \right)^{0.4}
\left( \frac {\eta} {0.3}  \right)^{-0.8}. \label{tmin}
\end{equation}

Soker \& Kastner (2003) estimate the total X-ray luminosity of the
shocked fast wind as
\begin {eqnarray}
L_x \simeq 10^{33}  
\left( \frac {T} {10^6 \K}  \right)^{-2} \left( \frac {v_{\rm c}}
{20 \km \s^{-1}} \right)^{-2} \left( \frac {t} {500 \yr}
\right)^{-2} \left( \frac {\Delta t} {100 \yr}  \right)
\label{lx1} \\ \nonumber \left( \frac {\dot M_2 v_2} {3 \times
10^{-5} M_\odot \yr^{-1} \km \s^{-1}}
   \right)^{2}
\left( \frac {\eta} {0.3}  \right)^{-2} \erg \s^{-1},
\end{eqnarray}
where $\Delta t$ is the time during which the central star blows
the fast wind segment at a speed of $v_2$ which is responsible for
most of the X-ray emission.

In the present work, we use the cooling function $\Lambda$ for
solar abundances from Sutherland \& Dopita (1993; their table 6),
linearly interpolating their data to obtain a continuous function
$\Lambda(T)$. Rapidly cooling regions having $\tau_{\rm cool}<t$
are not included in the computation of the X-ray emission. It is
important to note that this is not a fully self-consistent
treatment, because ($i$) we do not follow the evolution of gas
parcels with radiative cooling, but rather examine the radiative
cooling time at a specific time, ($ii$) we do not take into
account the process by which ambient gas from hotter regions of
the bubble fills in for the gas that has cooled to $T \ll 10^6
\K$, compressed, and now occupies only a small volume of the flow.
However, in the relevant self similar solutions studied here, we
find that only small regions of the flow are associated with short
cooling times $\tau_{\rm cool}<t$. Therefore, the present
treatment is in fact appropriate for the present purpose of
studying the behavior of the self-similar regime.

\subsection{X-Ray Emission}

The X-ray luminosity is calculated by integrating the emissivity
over the hot bubble .
\begin{equation}
L_x(R_c)=\int_{R_2}^{R_c} n_e n_p \Lambda_{0210} (T) 4 \pi r^2 dr,
\label{lx2}
\end{equation}
(but ignoring regions with $\tau_{\rm cool}<t$) where
$\Lambda_{0210}(T) n_e n_p$ is the power emitted per unit volume
in the energy range $0.2-10 \kev$. This range is chosen to reflect
approximately the sensitivity regime of the {\it Chandra} and {\it
XMM-Newton} telescopes and science instruments. The function
$\Lambda_{0210}(T)$, which serves here essentially as an X-ray
energy emission rate coefficient, is obtained from the APEC data
base (Smith et~al. 2001) using XSPEC version 11.3.1 (Arnaud 1996).

In most X-ray observations of PNs the spatial resolution is not
high, and/or there are not enough photons to obtain a spatially
resolved temperature profile. In those cases, the total X-ray
spectrum of the PN is fitted with an isothermal model, thus
assigning a single, best-fit temperature to the entire hot bubble.
For comparison with observations we therefore define an X-ray
weighted average temperature for the hot bubble by:

\begin{equation}
T_x= \frac{1}{L_x}\int_{R_2}^{R_c} T n_e n_p \Lambda_{0210}(T) 4
\pi r^2 dr. \label{tb1}
\end{equation}

\noindent or in terms of the more easily observed quantity, the
emission measure distribution $EMD(T)$

\begin{equation}
T_x= \frac{1}{L_x}\int_{R_2}^{R_c} T \Lambda_{0210}(T) EMD (T) dT.
\label{tb2}
\end{equation}

\noindent where $L_x$ is taken from Eq.~(7) and $EMD(T)$ is
defined as:

\begin{equation}
EMD(T)\equiv n_e n_p \frac{dV}{dT} = n_e n_p 4 \pi r^2
\frac{dr}{dT} \label{em}
\end{equation}

\noindent The right hand side of Eq. (10) is valid for the
spherically symmetric case. $EMD(T)$ can be calculated from the
self similar temperature profile $T(r)$ and the results will be
presented in the next section. Note that as for $L_x$, in
calculating $T_x$ cooling is taken into account by the exclusion
of fast cooling regions ($\tau_{\rm cool}<t$).  In practice, the
observed temperature is obtained by fitting spectral models to the
X-ray {\it photon} (not energy) spectrum using $\chi ^2$
minimization techniques (Arnaud ~1996). To that end, it would have
been more rigorous to use the X-ray {\it photon} emission rate
coefficient instead of $\Lambda_{0210}(T)$. In a sense it would
have also been better to take the square of $EMD(T)$ as the
distribution function in Eq. (9) instead of simply $EMD(T)$.
However, we have checked and found that these corrections have a
negligible effect of only a few percent on the calculated result
for $T_x$ (due to the dominance of the high-density
low-temperature regions in the flow). Therefore, for simplicity we
have used Eqs.~(8) and (9) in the forms given above.

\subsection{Outline of Calculations}

The set up of the calculations is as follows:
\begin{enumerate}
\item The parameters specified are: The time of the observation
$t$;  The velocities, $v_1$ and $v_2$ and the mass loss rates
$\dot M_1$ and $\dot M_2$, of the slow and fast winds,
respectively. \item The contact discontinuity velocity $v_c$ is
calculated from equation (\ref{vc}). \item The position (radius)
of the contact discontinuity is calculated by $R_c=v_ct$. \item
The position of the inner (reverse) shock $R_2$ (or of $\eta$) is
given by equation (\ref{r2}). \item The density $n(r)$ and
temperature $T(r)$ profiles in the hot bubble are found from the
self-similar solution of CI83; the self-similar equations are
integrated from $R_2$ to $R_c$, where at $R_c$ the gas follows the
contact discontinuity, i.e., $v=v_c$ ($U=1$ in the terminology of
CI83). \item The radiative cooling time $\tau_{\rm cool}(r)$ is
calculated at each radius by equation~(3), with $\Lambda(T)$ from
Sutherland \& Dopita (1993) and using $k_{cool}$~= 5/2 as
explained above. \item If $\tau_{\rm cool}<t$, the gas at that
radius is not included in the computation of the X-ray emission.
\item The X-ray luminosity and average temperature are calculated
from Eqs.~(7) and (8), respectively. \item The calculations are
repeated for different sets of parameters.
\end{enumerate}

\section{RESULTS}
In order to explore the self similar solutions to our problem we
carried out several runs, which are summarized in Table 1. We
examine three cases for the slow wind parameters: $\dot M_1=3
\times 10^{-6} M_\odot \yr^{-1}$, $\dot M_1=7 \times 10^{-6}
M_\odot \yr^{-1}$, and $\dot M_1=3 \times 10^{-5} M_\odot
\yr^{-1}$. For each $\dot M_1$ value, we take three $v_2$ values
of 300~\kms, 500~\kms, and 700~\kms. The slow wind velocity $v_1$
is assumed to be 10~\kms\ throughout our calculations. For all
runs $\dot M_2$ is chosen so that $\dot M_1 v_1 = \dot M_2 v_2$.
For each time $t$, we can calculate the temperature and density
profiles.

One property of the solution should be emphasized. The X-ray
emitting gas resides close to the contact discontinuity. Faster
gas with $v_2 \ga 700 \km \s^{-1}$, which was expelled more
recently in the evolution, does not contribute to the X-ray
luminosity. Therefore, it does not matter whether at later times
the mass loss rate and fast wind velocity were constant as assumed
here (this is the limitation of the self similar solution), or
whether the velocity increases and mass loss decreases as is known
to occur in PNs. In either case, the X-ray emission will come from
gas residing near the contact discontinuity that was expelled when
the PN was young, and even as early as the post-AGB phase when the
fast wind speed was $v_2 \simeq 400-600 \km \s^{-1}$. It is
expected that at later times the observed fast wind will be much
faster, as observed in many PNs. {\it In other words, it is not
the presently or recently blown wind that is responsible for the
X-ray emission.} This is why the self-similar solution is adequate
for our goal.

\begin{table}

Table 1: Cases Calculated

\bigskip
\begin{tabular}{|l|c|c|c|c|c|c|}
\hline
Run & $\dot M_1$ & $\dot M_2$   & $v_2$ & $v_c$  & $\eta$ &line \\
&$M_\odot \yr^{-1}$ & $M_\odot \yr^{-1}$ &$\km \s^{-1}$ &
$\km \s^{-1}$ & & \\
\hline
A3 &$3 \times 10^{-6}$ &$1\times 10^{-7}$ &$300$ &$28.6$ &0.52&solid  \\
\hline
A5 &$3 \times 10^{-6}$ &$6\times 10^{-8}$ &$500$ & $32.6$ &0.42 &dashed \\
\hline
A7 &$3 \times 10^{-6}$ &$4.33\times 10^{-8}$ &$700$ & $35.6$ & 0.37&dotted \\
\hline
B3 &$7 \times 10^{-6}$ &$2.33 \times 10^{-7}$ &$300$ &$28.6$ &0.52 & solid  \\
\hline
B4 &$7 \times 10^{-6}$ &$1.75 \times 10^{-7}$ &$400$ &$30.8$ &0.44 &   \\
\hline
B5 &$7 \times 10^{-6}$ &$1.40 \times 10^{-7}$ &$500$ & $32.6$ & 0.42& dashed \\
\hline
B6 &$7 \times 10^{-6}$ &$1.17 \times 10^{-7}$ &$600$ &$34.2$ &0.38 &   \\
\hline
B7 &$7 \times 10^{-6}$ &$ 1 \times 10^{-7}$ &$700$ & $35.6$ &0.37 &dotted\\
\hline
C3 &$3 \times 10^{-5}$ &$1 \times 10^{-6}$ &$300$ &$28.6$ &0.52 & solid  \\
\hline
C5 &$3 \times 10^{-5}$ &$6 \times 10^{-7}$ &$500$ & $32.6$ &0.42 &dashed  \\
\hline
C7 &$3 \times 10^{-5}$ &$4.28 \times 10^{-7}$ &$700$ & $35.6$ &0.37 &dotted \\
\hline
\end{tabular}

\footnotesize
\bigskip

Notes: (1) For all runs $v_1=10 \km \s^{-1}$ (2) $\eta \equiv
R_2/R_c$. (3) Runs B4 and B6 are just shown in Fig.~6 \normalsize
\end{table}

\subsection{General Characteristics}

The self similar solutions all have qualitatively similar
profiles, a good example of which are those shown in Fig.~2 and~3.
In Fig.~2 we have chosen $R_c = 3 \times 10^{16} \cm$ which
corresponds to $t= 291 \yr$ for Run B5. It can be seen that from
the inner radius $R_2$, the density rises outwards, first
gradually, but then sharply peaking (tending to infinity) as the
fast wind accumulates towards the contact discontinuity $R_c$. On
the other hand, the temperature is high around the inner radius
(where the density is low) and decreases to very low temperatures,
much below X-ray temperatures, towards $R_c$, which is expected at
these high densities at the infinite Mach number limit. Indeed, as
seen from Fig.~2, the pressure profile does not change
significantly within the hot bubble. The earlier a gas segment is
shocked, the closer it is to the present-day contact
discontinuity. Gas segments which were shocked early suffer large
volume expansion as the bubble grows, hence strong adiabatic
cooling. The mass profile (integral of mass density over volume)
is also plotted in Fig.~2. We have used this mass profile to
verify mass conservation in our calculations in the sense of:

\begin{equation}
\int_{R_2}^{R_c} \mu m_p n 4\pi r^2 dr = \dot M_2 t \label{eq10}
\end{equation}

\noindent where $\mu m_p$ is the average particle mass.

\subsection{Cooling Effect and X-Ray Emission}

The high-density outer region of the hot bubble can cool
relatively quickly. In particular, the cooling time can be shorter
than the age of the flow. Obviously, this effect becomes
increasingly important as the density increases with $\dot M_2$.
Fig.~3 presents the results for Run C5, again for $R_c = 3\times
10^{16} \cm$, which corresponds here to an age of $t$~= 291 \yr.
In this run, $\dot M_2=6 \times 10^{-7} M_\odot \yr^{-1}$, which
is about four times larger than $\dot M_2$ in Run B5. This is
reflected in the four-times higher density values seen in Fig.~3
compared with those of Fig.~2. The high density results in shorter
cooling times for Run C5. Consequently, larger regions of the hot
bubble near the contact discontinuity need to be disregarded when
calculating the X-ray emission.

\begin{figure}  
\vskip -2.2 cm
\resizebox{0.95\textwidth}{!}{\includegraphics{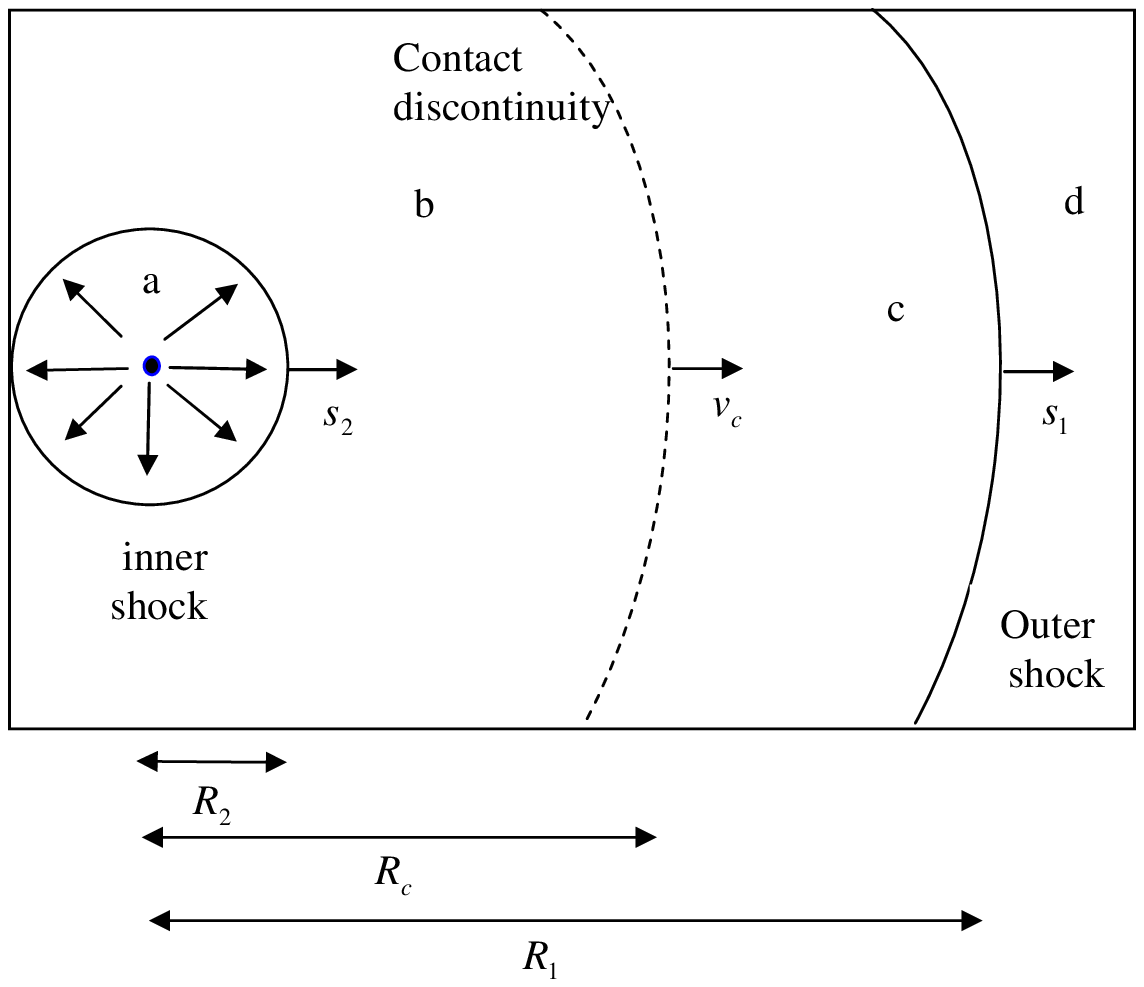}}
\vskip -2.2 cm
\caption{ {{{ A schematic drawing of the interacting stellar wind
flow structure. The regions indicated are: (a) the undisturbed
central star fast wind; (b) the hot bubble formed by the shocked
central star fast wind; (c) the shocked slow wind; and (d) the
nebular gas (the pre-shock AGB wind). $s_1$ and $s_2$ represent the
forward and reverse shocks velocities.}}} }
\end{figure}
\begin{figure}  
\resizebox{0.95\textwidth}{!}{\includegraphics{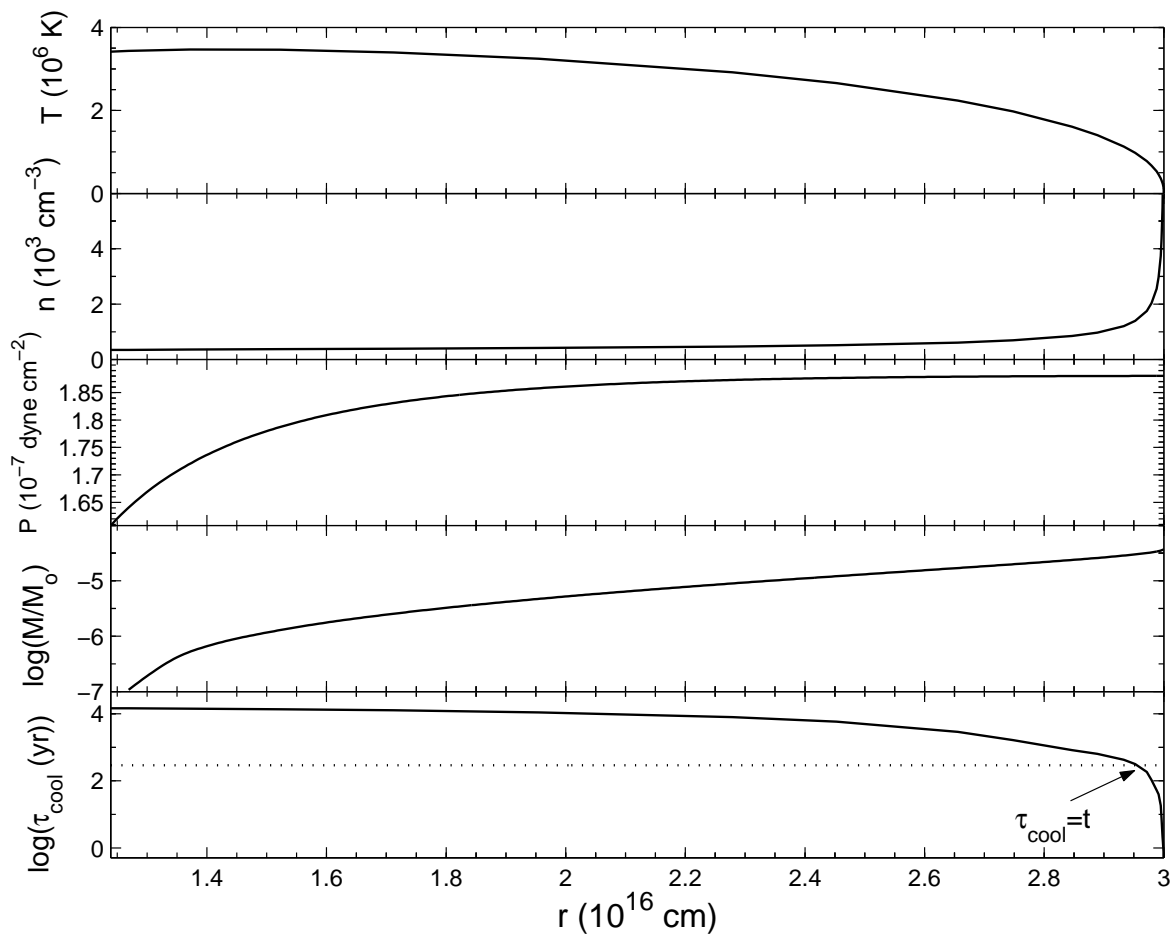}}
\caption{The gas temperature, total number density, pressure, the
total gas mass $M(r)$ (logarithmic scale in units of $M_\odot$),
and cooling time (logarithmic scale in years), between $R_2$ and
$R_c$ as a function of the radius $r$, for Run B5 at an age of
291~yr. The dotted line in the third panel mark the age of the
flow (the time since the beginning of the fast wind).}
\end{figure}
\begin{figure}  
\resizebox{0.95\textwidth}{!}{\includegraphics{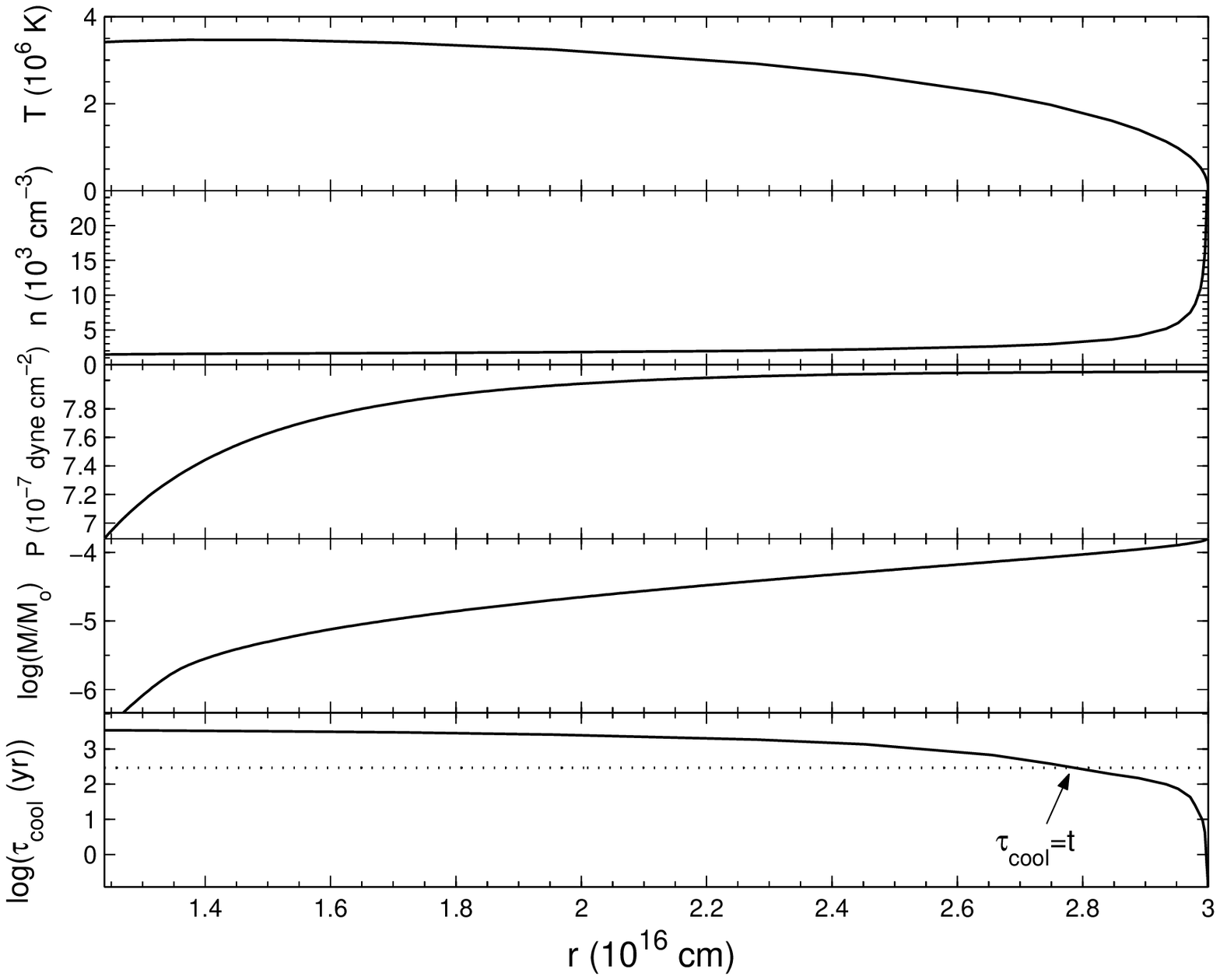}}
\caption{The same as Fig.~2 but for Run C5. The radius where
cooling time equals the age, is smaller than in the lower density
runs in Fig.~2. Note the different scales compared with Fig.~2. }
\end{figure}

In order to elucidate on the effect of the removal of the cool gas
on the total X-ray luminosity $L_x$ in the $0.2-10 \kev$ band, we
have plotted in Fig.~4 the total X-ray luminosity as a function of
time $t$, once taking radiative cooling into account (by
disregarding regions that have cooled- thick lines in Fig.~4), and
also when cooling is neglected (thin lines). The three cases
considered in Fig.~4, namely Runs B3, B5, and B7, are all for
$\dot M_1=7 \times 10^{-6} M_\odot \yr^{-1}$, but for different
fast-wind velocities.  As can be expected, in the slower wind case
(Run B3), the temperature is lowest and $\dot M_2$ is highest,
therefore the density is highest and the cooling time is shortest.
It can be seen that the effect of cooling is, indeed, most
significant for this case, whereas for higher velocities, the
cooling effect is essentially unimportant.

Fig.~4 also demonstrates that the X-ray luminosity $L_x$ is
highest for a fast wind velocity $v_2$~= 500~\kms, i.e., Run B5.
Both lower and higher velocities result in less X-ray emission
than in this case (B5). The lower luminosity for $v_2$~= 300~\kms\
 (B3) is due to the fact that the shocked gas is colder than
typical X-ray temperatures and a large fraction of the radiation
is emitted below $0.2~\kev$, which is our chosen lower limit for
$L_x$. The reason for the lower luminosity in the case of $v_2$~=
700~\kms\ (B7) is different. In this case, the temperature is high
and $\dot M_2$ is low, therefore the density is low.

In Fig.~4, we also plot $L_x$ as given by equation (\ref{lx1}).
That equation is derived under different assumptions than the ones
used in the self similar analysis. In order to make the comparison
meaningful, we take $\Delta t = t/5$ as in Soker \& Kastner
(2003), and the bubble temperature is taken to be the post-shock
temperature for Run B5, $T=3.5 \times 10^{6}$~K. The $t^{-1}$
dependence is roughly the same and the difference in X-ray
luminosity is relatively small between that found in run B5 and
that given by equation (\ref{lx1}), considering the different
assumptions entering the two methods (Soker \& Kastner 2003).
\begin{figure}[h]  
\resizebox{0.95\textwidth}{!}{\includegraphics{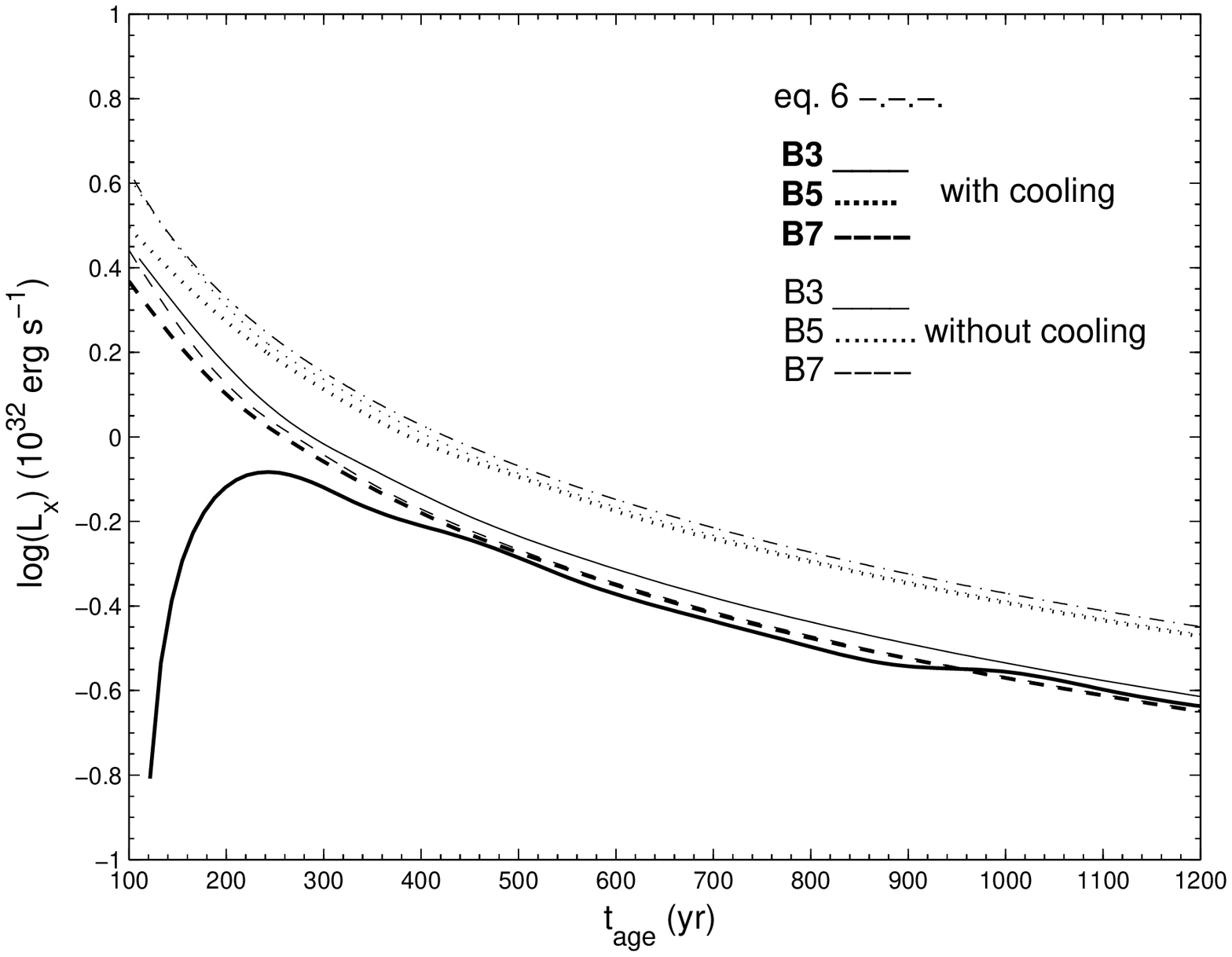}}
\caption{The luminosity as a function of the wind age
$t={R_c}/{v_c}$, for Runs B3 (solid lines) B5 ( dotted lines) and
B7 (dashed lines). Thick lines give the luminosity when gas
parcels with cooling time shorter than the bubble's age are not
considered; this will be the cases presented for the rest of the
paper. For comparison thin lines present the same run but when
cooling is not taken into account, and the X-ray contribution of
cold parcels of gas is included. The dashed-dotted line is $L_x$
as given in equation (\ref{lx1}) for the parameters of Run B5 from
Table 1, $\Delta t=t/5$, and $T=3.5 \times 10^6 \K$ (see text). }
\end{figure}

\section{COMPARISON WITH OBSERVATIONS}

\subsection {Luminosity}
The X-ray luminosity of the different Runs as a function of the
bubble age is given in Fig.~5. The plots are grouped according to
their $\dot M_1$ value (A, B, and C). The triangles represent the
location on the plot of the first five observed PNs summarized in
Table~2. The PNs He 3-1475 (Sahai et al.\ 2003) and Mz 3 (Kastner
et al.\ 2003) are not shown, nor are they listed in Table~2, as
their X-ray emission most likely comes from jets. From Fig.~5, it
appears that Run B5 can match the observations of the old PNs ($t
\ge$ 1000 yr) NGC~6543, NGC~7009, and NGC~2392, while Run C3
matches the observations of the younger PNs ($t <$ 1000 yr)
NGC~7027 and BD~+30~3639. The A runs clearly are not sufficiently
massive to produce the high luminosites observed. This indicates
that the observed PNs correspond to the higher $\dot M_1$ and
$\dot M_2$ values in our models ($\dot M_2$ up to $10^{-6} M_\odot
\yr^{-1}$). However, note that the results in Fig.~5 do not take
the measured X-ray temperature into account. As shown below, the
velocity of 300~\kms\ is too low to explain the high observed
temperatures. Consequently, we believe that Runs B at velocities
of 400 to 600~\kms\ best describe the physical conditions of the
X-ray bright PNs in the sample. Obviously, the observations are
biased towards the brighter X-ray PNs and do not detect the
fainter X-rays from less massive fast winds as efficiently.
\begin{figure}[h]  
\resizebox{0.95\textwidth}{!}{\includegraphics{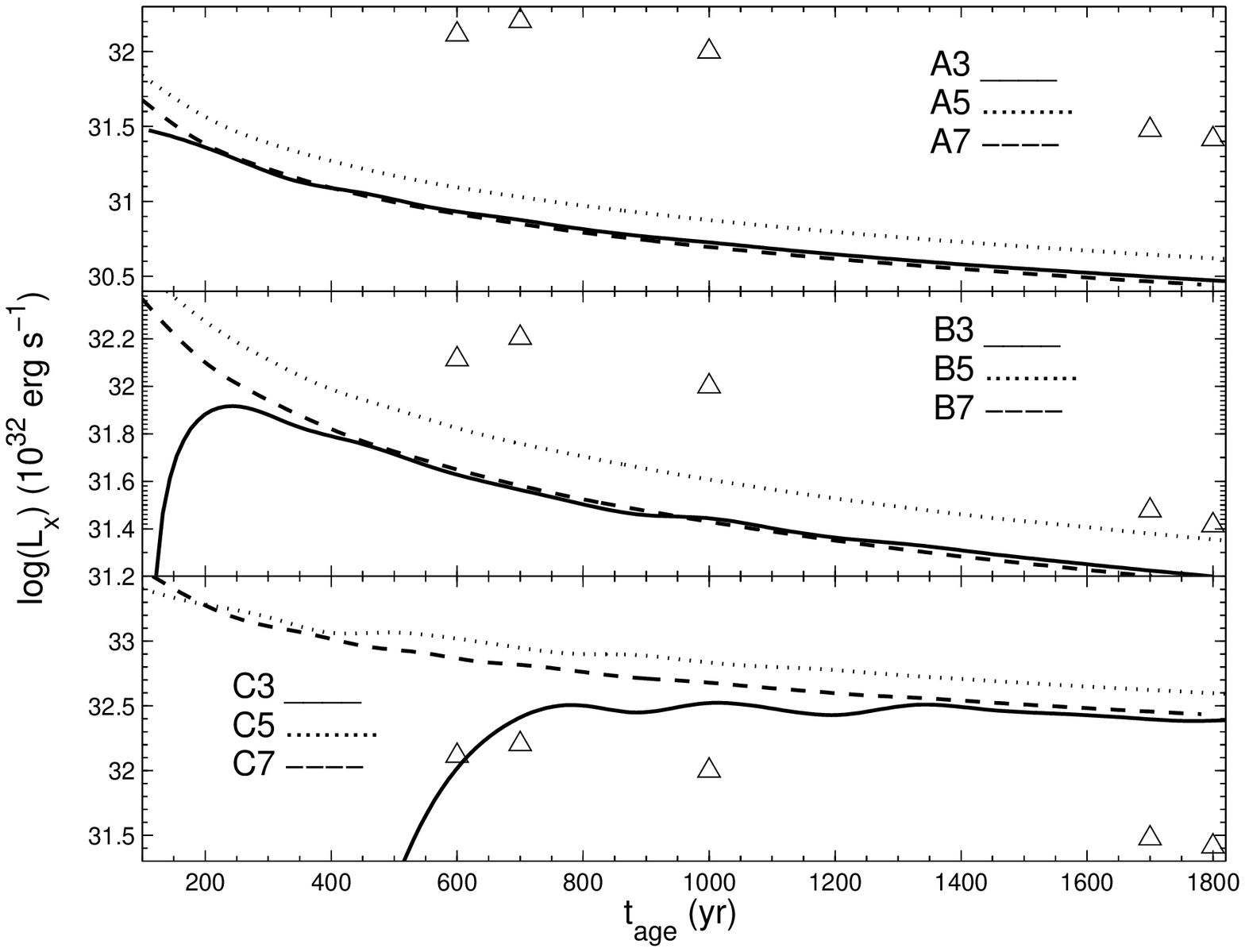}}
\caption{The luminosity as a function of the wind age
$t={R_c}/{v_c}$, for all cases. The triangles mark the position on
these plots of the five PNs listed in Table 2 (not including
NGC~40). Note the different scales in the three panels.}
\end{figure}


\begin{table}

Table 2: X-ray properties of Planetary Nebulae

\bigskip
\begin{tabular}{|l|c|c|c|c|}
\hline
\#&PN & $L_x$ & $T_x$ & Dynamical Age \\
 &  & $ 10^{32}$ erg s$^{-1}$ & $ 10^{6}$ K & yr\\
\hline
1& NGC 7027(PN G084.9-03.4) &1.3 &3 &600 \\
\hline
2& BD +30 3639(PN G064.7+05.0) &1.6 &3&700 \\
\hline
3&NGC 6543(PN G096.4+29.9) &1.0&1.7 &1000  \\
\hline
4 &NGC 7009(PN G037.7-34.5) &0.3 &1.8 &1700\\
\hline
5 &NGC 2392 (PN G197.8+17.3) &0.26 &2 & 1800\\
\hline
6 &NGC 40 (PN G120.0+09.8) &0.024 &1.5 & 5000\\
\hline
\end{tabular}

\footnotesize
\bigskip
The parameters of the first four PNs are summarized by Soker \&
Kastner (2003). The data for NGC 2392 are from Guerrero at al.\
(2005), and for NGC 40 from Kastner et al.\ (2005). \normalsize
\end{table}

\subsection{Temperature}
In order to further compare our results with observed PNs, we
follow the evolution with time of the different runs in the
temperature-luminosity plane. This enables us to analyze the three
central, observed parameters: $L_x$, $T_x$, and $t$
simultaneously. The average temperature of the bubble $T_x$ at
each time is calculated by equation (\ref{tb1}) and the evolution
of $L_x$ and $T_x$ are plotted in Fig.~6 for the various Runs. In
a self-similar flow when cooling is neglected, the temperature of
the bubble does not depend on time. However, in the present model,
which includes the cooling effect, the average temperature does
vary. At early times, when density is high and radiative cooling
times are still short in the low-T regions of the flow (eq.
\ref{tmin}), we do not include these regions for the computation
of the average temperature. Consequently, the calculated average
temperature at the early stages is higher than it is in the
late-time self similar solutions. As the bubble expands and the
density drops, cooling becomes less important and the temperature
tends to the self similar solution as seen in Fig.~6.

It can be seen from Fig.~6 that the evolution tracks of the B Runs
are most consistent with the observed PNs. This indicates that
these Runs can reproduce the observed luminosities and
temperatures. Out of the six PNs in the sample, three seem to lie
between the B4 and B5 Run indicating a fast wind velocity of
$v_2$~= 450--500~\kms, while the two youngest PNs are more
consistent with a fast wind velocity of $v_2$~= 600~\kms. The
oldest PN, NGC 40, is fitted better with even a slower fast wind,
having a velocity of $\sim 450 \km \s^{-1}$. This tendency of
older PNs to be better fitted with slower fast winds indicates
that the evolution with time of the properties of the fast wind
blown by the central star should be considered. Such an evolution,
namely, the increase in the fast wind speed and decrease of mass
loss rate with time (e.g., Steffen et al.\ 1998; Perinotto et al.
2004), cannot be considered with a self similar solution.
\begin{figure}   
\resizebox{0.95\textwidth}{!}{\includegraphics{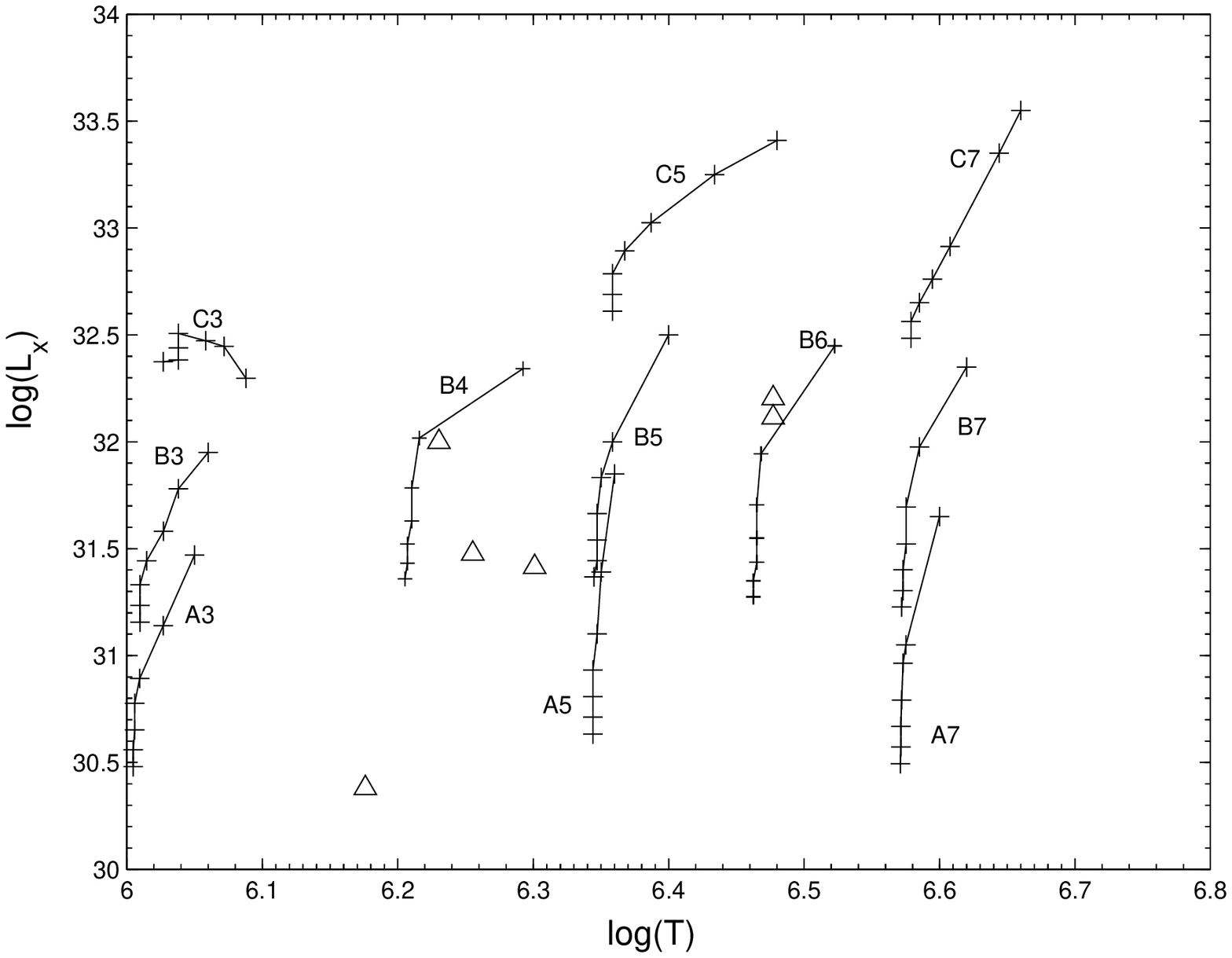}}
\caption{The evolution of the different cases with time in the
temperature-luminosity plane. The $+$ signs mark radii along the
track, from upper right to lower left, and in units of $10^{16}
\cm$: 1, 4, 7, 10, 13, 16, and 19. The triangles mark the position
of the six PNs given in Table 2 on these plots.}
\end{figure}
\begin{figure}   
\resizebox{0.95\textwidth}{!}{\includegraphics{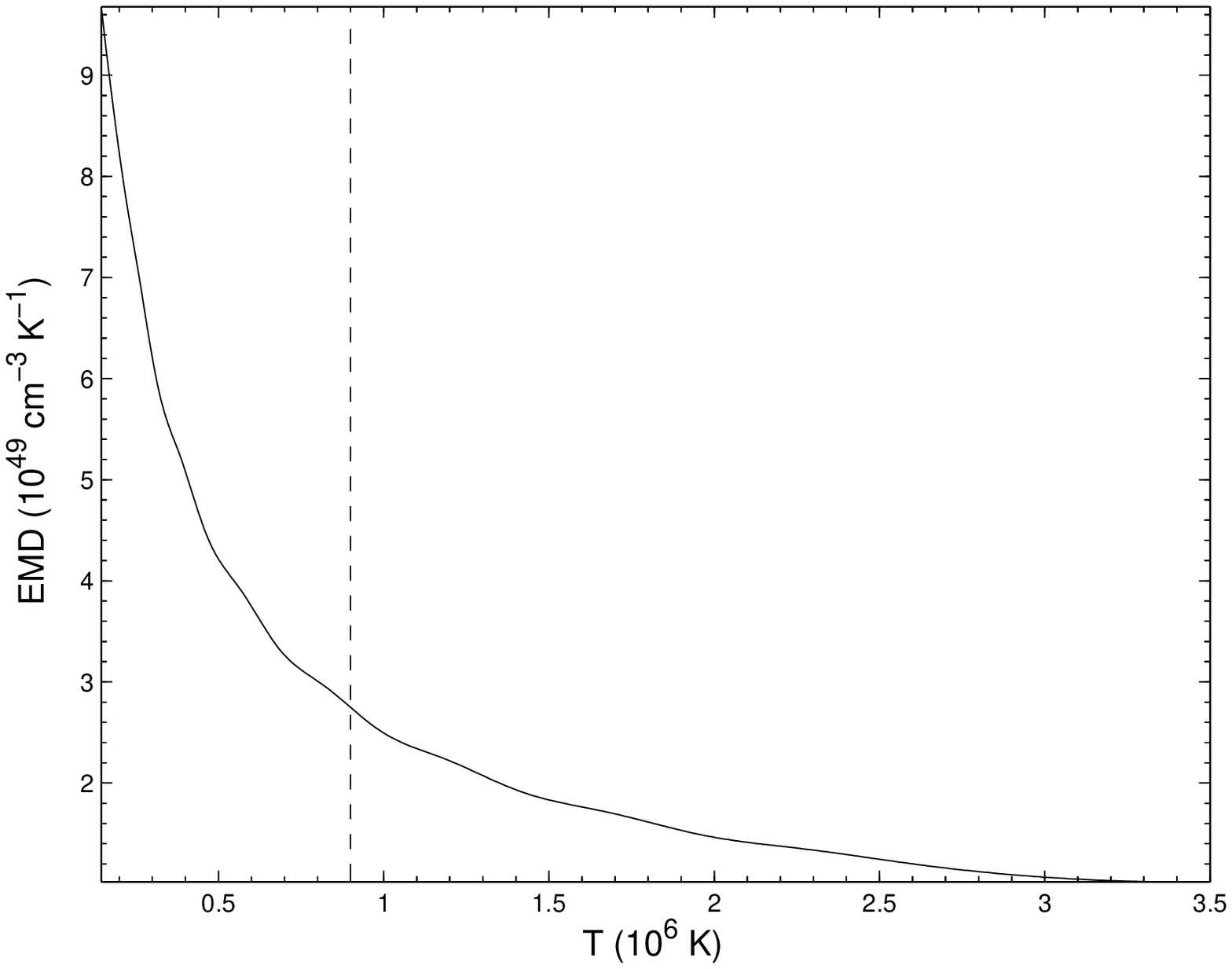}}
\caption{The emission measure distribution as a function of
temperature for run B5 at the age 291 yr. The dashed vertical line
is the cutoff for the cooling effect.}
\end{figure}

With the prospect of high spectral resolution X-ray observations
of PNs, there is interest not only in the average temperature
$T_x$, but also in the temperature distribution of the X-ray
source, i.e., in the explicit form of $EMD(T)$ (see Eq.~10). This
is the quantity that determines the emission-line details of the
X-ray spectrum (along with the elemental abundances). From the
self similar solution we have the explicit profile of $T(r)$,
which can be easily transformed into $EMD(T)$. In Fig.~7, we plot
$EMD(T)$ for the B5 Run at the age of 291 \yr. It can be seen that
$EMD(T)$ decreases sharply with temperature. This is why the
observed (and calculated) $T_x$ values are all relatively low,
only a few times 10$^6$~K. Most of the gas in terms of emission
measure is at low temperatures. The cutoff at low temperatures
observed in Fig.~7 is somewhat arbitrary and is due to cooling.
The lowest temperature gas cools rapidly below X-ray temperatures.
Of course, in a more realistic, full hydrodynamic model this
cutoff will be much more gradual. However, the effect of $EMD(T)$
peaking at low temperature $\sim$~10$^6$~K is expected to stay.
The explicit forms of $EMD(T)$ calculated from the self similar
solution can be used to predict the high-resolution spectra
expected from future observations of PNs. No such spectrum is
available yet.

\subsection{Surface Brightness}
When spatial resolution is sufficiently high and the number of
photons is large enough, X-ray images of PNs can be obtained. In
that case the image provides the surface brightness profiles. The
normalized surface brightness is calculated from our models by
integrating over the X-ray luminosity along the line of sight,
chosen here as the $z$ direction, taken here at projected radius
$d$ from the center of the bubble (see eq.~12)
\begin{equation}
S(d)=  2 \int_{0}^{\sqrt{R_c^2-d^2}} n_e n_p
\Lambda_{0210}[T(z^2+d^2)]dz .
\end{equation}

The surface brightness profile as a function of projected radius
$d$ is drawn in Fig.~8 for the B3, B5, and B7 Runs at ages of
300~yr and 1000~yr. The surface brightness rises sharply towards
the contact discontinuity, but then drops rapidly just before
reaching it. This is due to the efficient cooling in the
high-density region right behind the contact discontinuity. The
result is a bright X-ray region, which forms a ring close, but not
touching, the contact discontinuity. In a full hydrodynamic
calculation, the strongly emitting X-ray ring will be wider and
the rise and drop of surface brightness will be more gradual than
in Fig.~8.
As was noted in the past, a bright X-ray ring is expected from a
shocked central fast wind, but not from two fast jets (or a CFW).
In NGC~40 an X-ray ring is observed, and a spherical wind seems to
be present, as suggested by Kastner et al. (2005).
\begin{figure}   
\resizebox{0.95\textwidth}{!}{\includegraphics{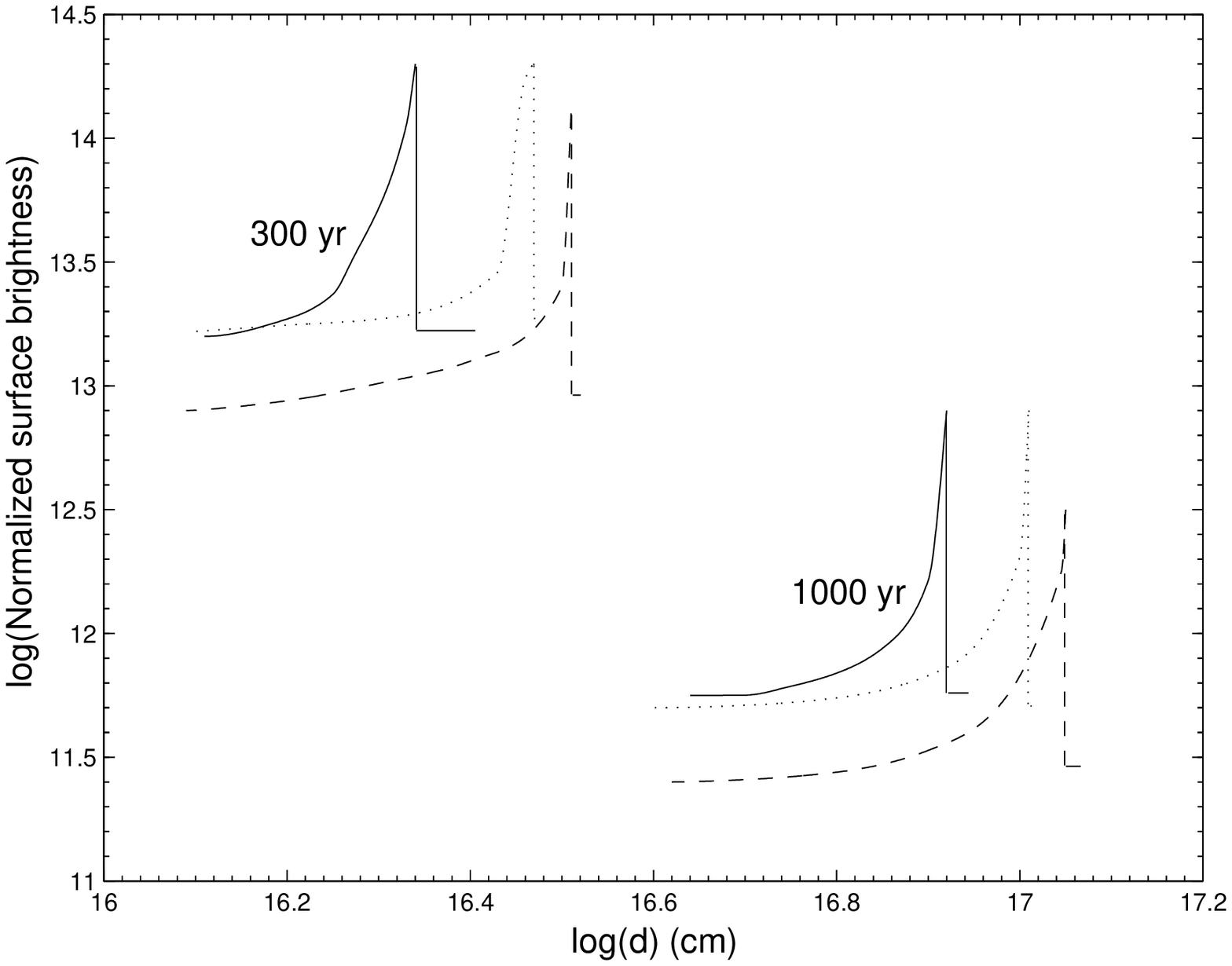}}
\caption{The surface brightness of the hot bubble after 300 year
(left) and after 1000 year (right) for Runs B3 (solid), B5
(dotted), and B7 (dashed). Plotted from $R_2$ to $R_c$ with cutoff
for $\tau_{\rm cool}<t$ .}
\end{figure}

{{{{
\subsection{Limitations of the Self Similar Approach}
The self similar approach has several well known limitations.
Here, we consider the two most important ones as far as the
X-ray emission is concerned.

\subsubsection {Time evolution}
The inability to incorporate time evolution of the fast wind is
indeed a limitation.
However, our results show that this should not have a serious impact on
the calculated X-ray flux and spectrum.
The X-rays emanate predominantly from up-stream wind segments
in the hot bubble, i.e., regions close behind the contact discontinuity
(Fig.~8), that were expelled early and over a relatively short time period
of $\sim 100-300\yr$.
The role of the wind blown later is merely that of a piston
exerting pressure on the hot bubble from behind, and thus
maintaining the high densities of the early-shocked gas.
The contribution of these late-time wind segments
to the X-ray emission is negligible as can be seen from Fig.~8.

In other words, according to our model, the details of the late-time evolution
of the fast wind are not important to the X-ray emission as long
as it continues to exert pressure on the high-density gas expelled
earlier.
Even if the gas in the inner regions of the hot bubble
originates from a faster (say even, $v_2 \sim \mathrm {few}~1000 \km \s^{-1}$)
and more tenuous wind, as expected for a more evolved post-AGB star,
the over all PN X-ray emission would remain more or less as predicted by the
self similar solutions, since the X-ray contribution of the tenuous, high-velocity
wind segments is considerably less than that of the early up--stream segments.

\subsubsection {Removing fast radiatively-cooling gas parcels}

Another limitation of the present self similar approach is the approximate
treatment of radiative cooling.
As discussed in section 2.2, in the calculation of the X-ray emission,
we did not include fast-cooling gas segments.
We considered only the X-ray emission from gas segments with
cooling times larger than the age of the nebula, namely, $\tau_{\rm
cool} > t$. This, of course, is not a fully consistent treatment.

In order to test the sensitivity of our results to this assumption, we
recalculated the X-ray evolutionary tracks of Fig.~6 for two cases,
but now varying the criteria for eliminating the fast cooling gas.
We chose the two runs, B4 and B6, most closely matching the observed
PNs in Fig.~6, and we varied the threshold for including gas segments in the
X-ray calculations from $0.25$ to $4$ times the age of the nebula.
This reflects more than an order of magnitude uncertainty in the relevant cooling
time (clearly more than enough). The results are plotted in Fig.~9.
As can be seen from Fig.~9, including gas with extremely
short cooling times (the lines marked $\tau_{\rm cool} >
t/4$, i.e. overestimating the amount of X-ray gas)
increases the calculated X-ray luminosity, while slightly
reducing the average X-ray temperature
(by a factor of $<10^{0.08}=1.2$).
These effects are noticeable only when the nebula is very young.
At higher PN ages the lines converge to the original $\tau_{\rm cool} =
t$ result.
Recall that the evolution in these tracks goes from high- to low-
temperatures, the first point in each track corresponding to a
contact discontinuity radius of $10^{16} \cm$ and an age of only
$\sim 100$~years.
Since most PNe are observed at much older ages, this result implies
that including fast-cooling gas does not have an appreciable effect on
our results for the X-ray luminosity nor the average temperature.

Taking our criterion to the other extreme and including in the
calculation of the X-ray emission only gas with very long cooling
times ($\tau_{\rm cool} > 2t, 3t, \mathrm {and}~4t$, i.e., thus
decreasing the amount of X-ray gas considered) produces
evolutionary tracks generally to the right (higher-T) of the
$\tau_{\rm cool} = t$ thick lines in Fig.~9. The result is to
gradually reduce the X-ray luminosity and to increase the average
X-ray temperature, as expected by eliminating more and more
high-density, low-T segments. Ignoring the first point in each
evolutionary track in Fig.~9 when the nebula is very young, the
luminosity drops by a factor of $<10^{0.3}=2$, and the temperature
rises by a factor of $<10^{0.08}=1.2$. This shows that even when
we disregard all of the gas with cooling times much larger, by up
to a factor of $4$, than the age of the nebula the results do not
change by more than a factor of 2 in the most important measurable
X-ray parameters. Only when we remove gas with cooling times as
long as 5 times the nebular age (essentially ignoring the entire
hot bubble) do we manage to reduce the X-ray luminosity much below
the typical observed values. Removing gas with such long cooling
times is clearly unjustified. Even removing gas with $\tau_{\rm
cool} \sim 3 t$ is not obviously justified.

To summarize this subsection, our main conclusion, that the
observed X-ray emission of a large fraction of PNs can be
accounted for by shocked wind segments that were expelled during
the early PN phase when the fast wind speed is moderate, $v_2 \sim$
400-600~\kms, and the mass loss rate is a few times $10^{-7}
M_\odot \yr^{-1}$ holds. Our approximate treatment of the evolution of the fast
wind and the exact radiative cooling scenario can not change these parameters by more than
a factor of $\sim 2$. In the future, we hope to be able to carry out numerical simulations
in order to study more closely the detailed processes responsible for the X-ray emission
 from PNe.

\begin{figure}   
\resizebox{0.95\textwidth}{!}{\includegraphics{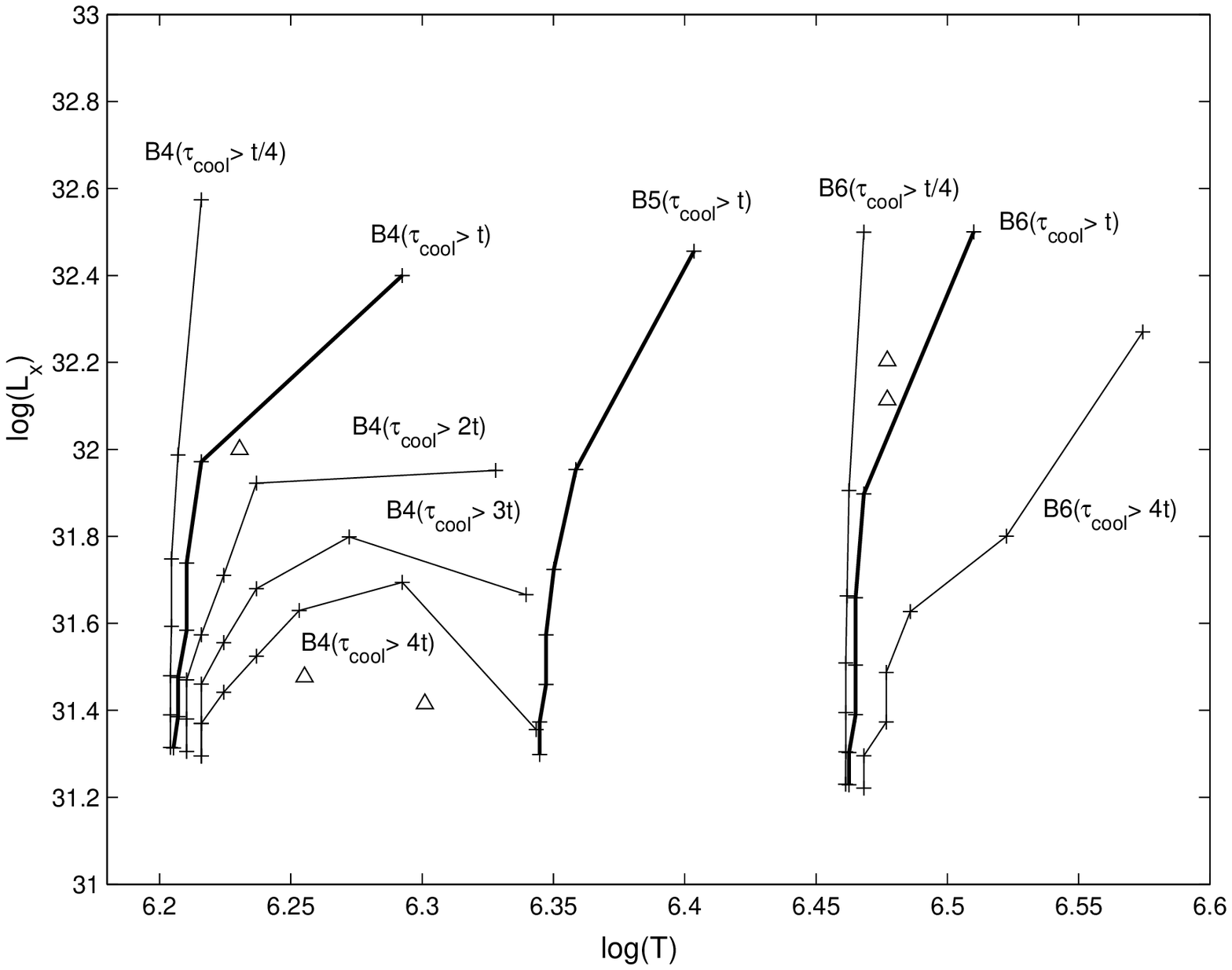}}
\caption{Evolution tracks as in Fig.~6, but only for models B4, B5, and B6, and here
we examine different criteria for removing cooled regions from the
computation of the X-ray emission. Thick lines show the results as
in Fig.~6, namely, only regions with cooling times $\tau_{\rm
cool}>t$ are included in the computation of the X-ray emission;
$\tau_{\rm cool}$ is the radiative cooling time (eq.
\ref{tcool1}), and $t(=R_c/v_c)$ is the age of the wind. The thin lines
show the evolution of runs B4 and B6 obtained using different criteria for removing
cooled regions. These are, as marked, $\tau_{\rm cool}>t/4$, $\tau_{\rm cool}> 2t$,
$\tau_{\rm cool}>3t$, and $\tau_{\rm cool}>4t$. }
\end{figure}
}}}}

\section{CONCLUSIONS}

We use the self similar solution for the collision of two
spherically symmetric concentric winds (CI83) to present a
quantitative treatment of the X-ray emission expected from the
shocked fast wind blown by the central stars of PNs and proto-PNs.
We compare our results with five X-ray bright PNs for which high
quality X-ray observations are available. The comparison
constrains the parameter space of the physical quantities relevant
to these PNs. {{{ With the assumptions made in our model, }}}
we find that the fast wind velocities of these PNs are between 400
and 600~\kms. Much higher or lower velocities can not produce the
high X-ray luminosities observed. Furthermore, we find that for
the observed PNs, the mass outflow rates in the fast wind {{{
under our assumptions, }}} is of the order of a few $10^{-7}
M_\odot \yr^{-1}$, perhaps even reaching $10^{-6} M_\odot
\yr^{-1}$. Our results strengthen the claim of Soker \& Kastner
(2003; see Fig.~4 here), that the strongest contribution to the
X-ray emission of the shocked fast wind blown by the central star
comes from shocked wind segments that were expelled during the
early PN phase when the wind speed was moderate $\sim 500 \km
\s^{-1}$. This is most clearly seen in Figures 5 and 6. As Soker
\& Kastner (2003), we also find that the fast wind can account for
the luminosity and temperature of the X-ray emitting gas in PNs.
However, the increase of the fast wind speed and decrease of the
mass loss rate must be considered to get a better match to
observations. This cannot be achieved with a self-similar
solution.

It has been suggested that the relative low X-ray temperature in
PNs results from heat conduction from the hot bubble to the cold
shell (Soker 1994), or from mixing of the hot bubble and the cold
shell material (Chu et al. 1997). The heat conduction process was
studied recently in great detail by Steffen et al. (2005), who
find that it can indeed account for X-ray properties of PNs. We do
not claim that heat conduction does not occur, or that it cannot
account for the X-ray properties of PNs. We rather argue that the
X-ray properties of PNs can be explained with the evolution of the
central star wind even if heat conduction is inhibited by magnetic
fields. Future observations and their comparison with models based
on heat conduction or wind evolution will determine the dominant
cause of the low temperature of X-ray emitting gas in PNs.

We have also calculated the expected X-ray surface brightness (in
relative units). In reality, where the fast wind evolves with
time, the X-ray bright ring will not be so thin, and the X-ray
emission will be spread on a wider ring. Still, the ring-shaped
X-ray emission is a clear signature that the X-ray emitting gas
comes from a spherically symmetric wind, and not from jets. This
has been suggested to be the case in NGC~40 (Kastner et al.\
2005). In some other PNs, such as He 3-1475 (Sahai et al.\ 2003)
and Mz 3 (Kastner et al.\ 2003), the morphology of the X-ray
emission clearly points at the presence of jets. This shows that
in different PNs there might be different origins for the X-ray
emitting gas, as noted by previous authors studying X-ray emission
from PNs, e.g., Guerrero et al.\ (2005). Therefore, the X-ray
emission can be used to shed light on the shaping process of PNs.

The self similar calculations includes the adiabatic cooling of
the gas, which must be considered for comparison with observations
(Soker \& Kastner 2003). However, it does not include the
radiative cooling. We have treated radiative cooling by not
considering shocked-gas shells with radiative cooling times
shorter than the age of the flow. The effect of removing fast
cooling gas segments has been shown to be important only for the
high-mass, low-velocity winds. The changing properties of the
blown fast wind with time, radiative cooling of the shocked fast
wind will be treated in a future paper where we will use a
numerical code.

\acknowledgments We thank Joel Kastner for helpful comments. This
research was supported in part by the Israel Science Foundation,
grant 28/03, and by the Asher Fund for Space Research at the
Technion. E.B. thanks the Israeli Army for its hospitality during
the last month of this project.

\end{document}